\documentstyle[12pt,aaspp4]{article}

\newcommand{\etal}{{\it et al.}\ }

\lefthead{von Hippel}
\righthead{Cluster White Dwarfs}

\begin{document}

\title{Contribution of White Dwarfs to Cluster Masses}

\author{Ted von Hippel}
\affil{Dept. of Astronomy, University of Wisconsin, Madison WI 53706 \\
email: ted@noao.edu}
\authoraddr{National Optical Astronomy Observatories, 950 N. Cherry Av.,
Tucson, AZ 85719}

\begin{abstract}

I have undertaken a literature search through 31 July 1997 of white dwarfs
(WDs) in open and globular clusters.  I have tried to make a careful
evaluation in each case of the likelihood that the object is a white dwarf
and that it is a cluster member.  The results are presented for 13 open
clusters and 11 globular clusters.  Currently there are 36 single WDs and
5 WDs in binaries known among the open clusters, and 340 single WDs and 11
WDs in binaries known among the globular clusters.  From these data I have
calculated WD mass fractions for four open clusters (the Pleiades, NGC
2168, NGC 3532, and the Hyades) and one globular cluster (NGC 6121).  I
develop a simple model of cluster evolution that incorporates stellar
evolution but not dynamical evolution to interpret the WD mass fractions.
I augment the results of my simple model by turning to sophisticated
N-body simulations incorporating stellar evolution (Terlevich 1987; de la
Feunte Marcos 1996; Vesperini \& Heggie 1997).  I find that even though
these clusters undergo a range of degrees of kinematical evolution from
moderate (the Pleiades, NGC 2168, and NGC 3532) to strong (the Hyades and
NGC 6121) the WD mass fraction is relatively insensitive to kinematical
evolution and little changed from a model incorporating only stellar
evolution with a Salpeter-like initial mass function.  By comparing the
cluster mass functions to that of the Galactic disk, and incorporating
plausibility arguments for the mass function of the Galactic halo, I
estimate the WD mass fraction in these two field populations.  I assume
the Galactic disk is $\sim10$ Gyrs old (Winget \etal 1987; Liebert, Dahn,
\& Monet 1988; Oswalt \etal 1996) and that the Galactic halo is $\sim12$
Gyrs old (Reid 1997b; Gratton \etal 1997; Chaboyer \etal 1998), although
the WD mass fraction is insensitive to age within this regime.  I find
that the Galactic halo should contain from $8$ to $9$\% ($\alpha = -2.35$)
or perhaps as much as $15$ to $17$\% ($\alpha = -2.0$) of its stellar mass
in the form of WDs.  The Galactic disk WD mass fraction should be $6$ to
$7$\% (for a median stellar age of $5$ to $7$ Gyrs and $\alpha = -2.35$),
consistent with the empirical estimates of $3$ to $7$\% (Liebert, Dahn, \&
Monet 1988; Oswalt \etal 1996).

\end{abstract}


\section{Introduction}

Since white dwarfs are faint for most of their evolutionary lifetime,
their mass fraction in clusters and in the field is difficult to measure.
Yet the WD mass fraction is important both for the dynamical evolution of
star clusters and potentially for the mass of the Galactic disk and halo.
Even in the immediate solar neighborhood, the range of the WD mass density
estimates vary by more than a factor of two, from $2.0 \times 10^{-3}
M_{\sun} \ pc^{-3}$ (Liebert, Dahn, \& Monet 1988) to $4.6^{+2.2}_{-0.4}
\times 10^{-3} M_{\sun} \ pc^{-3}$ (Oswalt \etal 1996).  While the solar
neighborhood stellar density itself is poorly constrained, for a value of
$\sim6.4 \times 10^{-2} M_{\sun} \ pc^{-3}$ (Mihalas \& Binney 1981; and
consistent with Kuijken \& Gilmore 1989, after subtracting the
interstellar gas mass) the WD mass fraction ranges from $3$ to $7$\%.  In
the Galactic halo the situation is even more poorly constrained, and the
WD mass fraction is effectively observationally unknown.  Indeed, studies
of gravitational lensing in the Milky Way (e.g.\ Alcock \etal 1997) led to
a flurry of papers during 1997 examining whether $\sim50$\% of the
Galactic dark matter could be in the form of halo WDs.  The bulk of these
studies concluded that such a high halo WD mass fraction can be ruled out
(see Gibson \& Mould 1997, and references therein), but the mere fact that
the mass fraction of WDs is so poorly known drives speculation about its
importance.  For the clusters, the presumed source of the field WDs, the
WD mass fraction must depend on the cluster age and kinematical evolution
(e.g.\ Vesperini \& Heggie 1997).  In the last three years a number of
studies have identified and measured the properties of WDs in open and
globular clusters.  Most of these new cluster WD measurements have been
made possible by the ability of the Hubble Space Telescope (HST) to detect
very faint point sources and separate them from the many faint resolved
background galaxies.  These studies have been motivated by the independent
information available from cluster WDs on cluster distances (Renzini \etal
1996), cluster ages (von Hippel, Gilmore, \& Jones 1995), and constraints
on stellar evolution (Richer \etal 1997).  An important byproduct of these
studies is the number and mass contribution of WDs to their parent
clusters.  To the best of my knowledge no one has yet extracted this
important information.  In this paper I first tabulate the known cluster
WDs and estimate their fraction by mass in a handful of clusters.  I then
use a simple interpretive model supplemented by cluster dynamical studies
in the literature to argue that the observed numbers of cluster WDs are
about what one would expect based on stellar evolution theory alone and
are insensitive to the cluster dynamical history.  Finally, I discuss the
relevance of the cluster WD mass fractions to the disk and halo field star
WD mass fractions.

\section{Survey of Observations}

Starting with the NASA ADS Abstract Service I performed a literature
search on white dwarfs in open and globular clusters through 31 July
1997.  I included cataclysmic variables and other types of binary systems
where the authors specifically discussed the WD nature of one of the
binary components.  My literature search covered 49 open cluster
references and 82 globular cluster references.  In assessing whether an
object was a cluster WD, I assessed the likelihood of cluster membership as
well as the likelihood that the object is a WD.  For the globular clusters
I required that the authors gave a high likelihood of the object being a
cluster member and being a WD, although most of the globular cluster WDs
were identified purely on the basis of multi-color photometry.  For the
globular cluster photometric candidates, I checked that they had the
appropriate colors and magnitudes for the cluster distances and that there
were few, or no, field stars with the same colors and magnitudes.
Nontheless, especially near the limit of the photometry, it is difficult
to judge the number of genuine WDs identified.  For the open clusters,
where field contamination is much more problematical, I was stricter about
membership probabilities and required that the authors used proper motions
or some other criteria to evaluate membership and that the resulting
membership probability was ``probable'' or better.

The results of the literature search for open and globular clusters are
presented in Tables 1 and 2, respectively.  In both tables columns 1 \& 2
list the names of the clusters, column 3 lists the numbers of known single
WDs, column 5 lists the numbers of known WDs in binaries, column 7 lists
the numbers of WD members calculated to exist, and column 8 lists the
total cluster masses in solar masses.  Columns 4 and 6 provide references
to the previous columns, whereas column 9 provides references to the
previous two columns.  Table 1 has two more columns than Table 2, and its
column 10 lists the cluster ages in Gyrs, with references in column 11.
The cluster masses are often lower limits and generally apply to cluster
stars within some luminosity or mass range and/or within some central area
of the cluster.  For the open cluster ages there were often multiple
references and I have chosen recent and representative values.
Nontheless, I represent the age range with the error bars in Figure 2.  I
assume that all globular clusters listed in Table 2 are $12 \pm 2.3$ Gyrs
old based on recent Hipparcos subdwarf studies (Reid 1997b; Gratton \etal
1997; Chaboyer \etal 1998).  This topic is discussed further, below,
although precise ages are not critical to the results of this paper.

Using the values listed in Tables 1 \& 2 I was able to estimate WD mass
fractions for four open clusters (the Pleiades, NGC 2168, NGC 3532, and
the Hyades) and one globular cluster (NGC 6121).  Following is a brief
discussion of how I arrived at each of the cluster WD mass fractions.  The
discussion is ordered by increasing cluster age.

\subsection{Open Clusters}

Pleiades: There is one known WD cluster member with a mass of $0.98
M_{\sun}$ (Bergeron, Saffer, \& Liebert 1992).  It is unlikely that there
are any undiscovered Pleiades WDs since the proximity and youth of the
Pleiades makes any cluster WD relatively bright and easy to detect.  It is
still possible, however, that one or two Pleiades WDs exist as close
companions to one of the brightest cluster stars.  The total cluster mass
is $1000$ to $2000 M_{\sun}$ (van Leeuwen 1980; Meusinger, Schilbach, \&
Souchay 1996).  Assuming the single known WD is the only cluster WD, the
Pleiades WD mass fraction is $7.4 \pm 2.5 \times 10^{-4}$.  The Pleiades
is 70 Myrs old (Stauffer, Hamilton, \& Probst 1994) with a main sequence
turn off mass of $5.3 M_{\sun}$ (Weidemann 1977).

NGC 2168 (= M35): There are two known WD members, each with masses of $0.7
\pm 0.1 M_{\sun}$ (Reimers \& Koester 1988a,b).  There are unlikely to be
other single cluster WDs, but the constraints on members of multiple
systems are weak.  The total cluster mass is at least $1600$ to $3200$
$M_{\sun}$ (Leonard \& Merritt 1989).  Since both the WD count and cluster
mass are lower limits, and both are unlikely to be more than a factor of 2
too low, I will assume that the ratio of the two is roughly correct.  The
WD mass fraction for this cluster is then $6.6 \pm 3.3 \times 10^{-4}$,
where I have increased the error estimate by $50$\% to reflect the
uncertainties inherent in the two lower limits.  NGC 2168 is $85 \pm 15$
Myrs old with a main sequence turn off mass of $\sim5 M_{\sun}$ (Reimers
\& Koester 1988a,b).

NGC 3532: There are six known cluster WDs with a total mass of $\sim4.6
M_{\sun}$ (Reimers \& Koester 1988a, 1989; Koester \& Reimers 1993).  As
is the case for NGC 2168, this is a lower limit due to possible WDs in
multiple systems.  It is also a lower limit in that only the central $30
\times 30$ arc minutes of the cluster have been surveyed for WDs.
Regardless, the total WD count is unlikely to more than double.  The total
cluster mass in the same central region is $\geq 600 M_{\sun}$ (Reimers \&
Koester 1989).  I believe this is a weaker constraint than the WD count,
and therefore the WD mass fraction is an upper limit of $\leq 7.7 \times
10^{-3}$.  NGC 3532 is $165 \pm 35$ Myrs old (Reimers \& Koester 1989)
with a main sequence turn off mass of $3.8 \pm 0.6 M_{\sun}$ (Reimers \&
Koester 1988a, 1989; Koester \& Reimers 1993).

Hyades: Despite the size of the Hyades on the sky, it is near enough and
its population has been well enough studied, that it is likely that all of
its WDs have been found.  This includes seven single WDs and three WDs in
binaries (HD 27483 consists of two F6 V stars and one WD, Boehm-Vitense
1993; HZ 9 consists of an M4.5e V star and a WD, and V471 Tau consists of
a K2 V star and a WD, White, Jackson, \& Kundu 1993).  The total mass in
these 10 Hyades WDs is $6.4 M_{\sun}$.  The expected error in the total
mass is smaller than the errors in the individual masses, which are
generally less than $5$\%.  There are currently three more candidate
cluster WDs, but they are unlikely to be members (Reid 1997a).  The total
mass of the Hyades was estimated by Reid (1992) to be $410$ to $480$ solar
masses.  The Hyades WD mass fraction is $1.45 \pm 0.15 \times 10^{-2}$.
The Hyades is $625 \pm 50$ Myrs (Perryman \etal 1998) old with a main
sequence turn off mass of $2.3 M_{\sun}$ (Boehm-Vitense 1993).

\subsection{Globular Clusters}

NGC 6121 (= M4): Because of the distance and age of NGC 6121 current
observations sample only the brighter portion of the WD cooling sequence,
with the faintest WDs expected at V $\geq 31$.  Additionally, to reach
even the brighter WDs in NGC 6121 requires the Hubble Space Telescope, and
so observations cover only a small part of the cluster field.  Although
the WD mass fraction cannot be estimated directly, as done (above) for
open clusters, it can still be derived by counting the number of
Horizontal branch stars and knowing their evolutionary lifetime in
comparison to the lifetime of the cluster WDs (essentially the lifetime of
the cluster).  Richer \etal (1995) used this technique and estimated that
the number of WDs expected in NGC 6121 is $2 \times 10^4$.  No error
estimates were given, so I assume an error of $\pm 1 \times 10^4$.  Among
the more than $200$ WDs that Richer \etal (1997) find in NGC 6121, they
estimate a mean mass of $0.51 \pm 0.03 M_{\sun}$.  Since the observable
(i.e.\ brighter) WDs are strongly weighted to those that have evolved off
the main sequence in the last $\sim5$ Gyrs, I make the small correction to
$0.55 M_{\sun}$ as the mean cluster WD mass.  Modern mass estimates for
NGC 6121 based on dynamical models range from $4.3 \times 10^4 M_{\sun}$
(Peterson, Rees, \& Cudworth 1995) to $\sim10^5 M_{\sun}$ (Sigurdsson
1993).  I take the mean of these two estimates and use the range as the
error estimate:  $M_{cluster} = 7.2 \pm 2.9 \times 10^4 M_{\sun}$.  The
NGC 6121 WD mass fraction is $0.15 \pm 0.10$.  For NGC 6121, as well as
the rest of the globular clusters listed in Table 2, I assume ages of $12
\pm 2.3$ Gyrs (Reid 1997b; Gratton \etal 1997; Chaboyer \etal 1998).
Although these ages are still a topic of debate, Figures 2 and 3
demonstrate that the WD mass fraction loses age sensitivity well before
$12$ Gyrs.  The main sequence turn off mass is globular clusters is
$\sim0.85 M_{\sun}$.

\section{A Simple Interpretive Model}

The WD mass fractions for the four open clusters and one globular cluster
support a general picture of an increasing WD mass fraction from $< 1$\%
at an age of approximately $100$ Myrs (Pleiades, NGC 2168, NGC 3532) to
$\sim1$\% by $1$ Gyr (Hyades) to $\sim15$\% by $\sim10$ Gyrs (NGC 6121).
How reasonable is such an interpretation?  To fully address this question
would require a comparison of the data to detailed cluster models that
fully incorporate stellar evolution and cluster dynamics.  Vesperini \&
Heggie (1997) have created just such model globular clusters using a
sophisticated N-body treatment incorporating stellar evolution and they
even explicitly followed the cluster WD mass fractions.  Terlevich (1987)
and de la Feunte Marcos (1996) used similar theoretical treatments to
investigate the general evolution of open clusters, though they did not
specifically investigate the evolution of the WD mass fraction.  Since the
currently available theoretical results do not cover the entire range of
cluster ages and physical parameters, I will tie together the open cluster
and globular cluster data with a simple interpretive model that
incorporates only the effects of stellar evolution and not dynamical
evolution.  To correct for the effects of dynamical evolution I will,
where possible, use the results of the above-mentioned theoretical
studies.

I assume all open and globular clusters were created in a single burst
star formation event and that their initial mass functions (IMFs) can be
characterized by a single or a double power law of the form

\begin{equation}
   N \sim M^{\alpha}, 
\end{equation}

\noindent
over the mass range $0.1 \leq M / M_{\sun} \leq 80$.  Current HST work on
globular clusters (Piotto 1997) that are thought to have suffered little
stellar evaporation is consistent with a single power law MF up to the
present-day turn-off mass, $\sim0.85 M_{\sun}$, at least at the precision
necessary for calculating WD numbers.  I consider a range of single power
law IMF slopes, $\alpha = 0, -1, -2, -2.35$, and $-3$, and one double
power law IMF slope, $\alpha = -2$ for $M \geq 0.6 M_{\sun}$ and $\alpha =
-1$ otherwise.  The double power law IMF is essentially the Galactic disk
MF given by Gould, Bahcall, \& Flynn (1997).  On this system, the slope of
$-2.35$ is the Salpeter (1955) value.  Stars evolve from the zero age main
sequence through the asymptotic giant branch on timescales given by the
stellar evolution parameterizations of Eggleton, Fitchett, \& Tout (1989)
and Tout \etal (1997).\footnote{Specifically, the main sequence, subgiant,
and red giant lifetimes are given by Equations A3, A11, and A19,
respectively, of Eggleton, Fitchett, \& Tout (1989).  The core He-burning
lifetime is given by Equations 6, A1, and A17 of Tout \etal (1997).} While
the parameterizations used here are all for solar metallicity stars, the
differential effect on the WD mass fractions is slight, with only a small
difference in the turn off mass as a function of metallicity affecting the
overall mass in main sequence stars.  White dwarfs are produced from post
asymptotic giant branch stars via the initial -- final mass relation.  I
have tried two different initial -- final mass relations; one given by von
Hippel, Bothun, \& Schommer (1997) based on the data compiled by Weidemann
\& Koester (1983),

\begin{equation}
   M_{WD} = 0.48 - 0.016 \ M_{ZAMS} + 0.016 \ M_{ZAMS}^2,
\end{equation}

\noindent
and the other the ``standard model'' parameterization of Wood (1992),

\begin{equation}
   M_{WD} = 0.49462 \ e^{(0.09468 \ M_{ZAMS})},
\end{equation}

\noindent
where $M_{ZAMS}$ is the zero-age main sequence mass and $M_{WD}$ is the
mass of the resulting white dwarf, both in solar masses.  Although these
two parameterizations are different, with the Wood standard model
parameterization being nearly a straight line, they yield essentially the
same results since the IMF and stellar evolution lifetimes are the main
determinants of the WD mass fractions.  This is encouraging since even if
the initial -- final mass relation is different at globular cluster
metallicities, it is unlikely to significantly alter the WD mass
fractions.

The highest mass main sequence star that forms a WD is most likely $\sim8
M_{\sun}$ (Koester \& Reimers 1996).  There is some question, however,
whether this upper mass limit varies, depending perhaps on stellar
abundances or rotation (Weidemann 1977), and it may be as low as $\sim5
M_{\sun}$ in some clusters.  Both of these upper mass limits are used in
this model.  All gas ejected from evolving stars and all neutron stars and
black hole remnants are assumed to leave the cluster.  Globular cluster
gas masses have been shown to be negligible (Krockenberger \& Grindlay
1995, and references therein), and the number of detected neutron stars is
small enough (e.g.\ Manchester \etal 1991) and neutron star kicks are
expected to be high enough (e.g.\ Helfand, Taylor, \& Manchester 1977),
that most neutron stars should leave the cluster.  The model does not
include binary stars.  Open clusters are known to have a large number of
binaries while globular clusters have binary fractions typically $\leq
5$\% (e.g.\ Richer \etal 1997).  The challenge in comparing this simple
model to the clusters is to observationally correct for binaries in the
open clusters.  It does not matter, for instance, that this model would
not predict cluster cataclysmic variables.  The key point is to predict
the expected mass fraction of WDs as a function of stellar population
age.  Finally, as discussed above, cluster kinematical evolution is not
incorporated.

Figure 1 shows the fraction of mass lost from the model clusters as a
function of age, up to $15$ Gyrs, for IMFs characterized by slopes $\alpha
= 0, -1, -2, -2.35$ (Salpeter 1955), and $-3$.  A double power-law slope
case is also plotted with $\alpha = -2$ above $0.6 M_{\sun}$ and $\alpha =
-1$ otherwise (as advocated Gould, Bahcall, \& Flynn (1997) for the
Galactic disk field stars).  The dashed lines are for model runs with
$M_{up} = 5 M_{\sun}$ and the solid lines are for model runs with $M_{up}
= 8 M_{\sun}$.  In this figure, the two different initial -- final mass
relations (Equations 2 and 3) would be indistinguishable, and so only
model runs based on the quadratic initial -- final mass relation (Equation
2) are plotted.  Clearly IMF slopes as flat as $0$ or $-1$ would cause the
cluster to evaporate (see also Terlevich 1987).  Even with slopes near the
Salpeter value, much of the initial cluster mass is lost and it is
essential to keep track of mass loss, since it is significant enough to
affect the total cluster mass, and hence any calculated WD mass
fractions.

Figure 2 shows the WD mass fraction for the model clusters as a function
of the IMF slope, for $M_{up} = 5$ and $8 M_{\sun}$.  Again the model runs
show only the results from the quadratic initial -- final mass relation as
the results for the exponential initial -- final mass relation differed by
only $0$ to $3$\%, depending on the IMF slope, $M_{up}$, and age.  (The
difference is always $\leq 1.2$\% for a Salpeter IMF slope.)  Also plotted
are the four open cluster and one globular cluster WD mass fractions,
along with their 1 $\sigma$ uncertainties.  In order to make the Hyades
and NGC 6121 data points visible, the $\alpha = 0$ and $-1$ model runs are
not plotted in their entirety.  Figure 3 is similar to Figure 2 except
that both axes are plotted as logarithms.  The model IMF slopes are the
same as in Figure 2.  The onset of WD creation for $M_{up} = 8$ near
Log(age) = $-1.3$ Gyrs and for $M_{up} = 5$ near Log(age) = $-0.7$ Gyrs
can be simply understood as the stellar evolutionary lifetimes for $8$ and
$5$ solar mass stars.  The arrow symbol near Log(age) = $-0.8$ Gyrs is the
upper limit value for NGC 3532.  It is clear from Figures 2 and 3 that the
WD mass fractions for the four open clusters and one globular cluster are
roughly consistent with an IMF with a Salpeter-like slope.  For the two
youngest open clusters, i.e.\ those with ages less than 100 Myrs, the WD
mass fractions display perhaps more sensitivity to the exact value of
$M_{up}$ than to the IMF slope.  Additionally, even if clusters IMFs can
be fit by power laws, the number of high mass stars is likely to be small
and should stochastically vary.

The WD mass fractions for a few representative old stellar populations are
listed in Table 3.  Column 1 lists the population age in Gyrs, and columns
2 through 5 list the WD mass fractions for four different IMF slope and
$M_{up}$ combinations, as labeled.  The WD mass fractions range from $6$\%
to $17$\% and are relatively insensitive to $M_{up}$ and age.  The primary
sensitivity at these ages is to the IMF slope.

\section{Discussion}

\subsection{The Role of Kinematical Evolution}

Kinematical evolution causes mass segregation and stellar evaporation.
Mass segregation alone is not expected to cause significant problems for
my simple model since observations of open clusters often cover the entire
cluster and observations of globular clusters are generally made near a
few core radii (e.g.\ De Marchi \& Paresce 1995a, 1995b) where King models
(e.g.\ Cool, Piotto, \& King 1996) and N-body simulations (Vesperini \&
Heggie 1997) have consistently shown that the present-day MFs (PDMFs) are
very similar to the global MF.  Generally, under a number of conditions
relevant to the distant globular clusters, these global MFs are very
similar to the IMF (Vesperini \& Heggie 1997).  For example, according to
Equation 16 of Vesperini \& Heggie (1997), even for a globular cluster
with $R_{peri} = 4$ kpc, $M_{initial} = 10^5 M_{\sun}$, and age $\approx
12$ Gyrs, an IMF slope $\alpha = -2.5$ population should be very similar
to the PDMF, which would have $\alpha = -2.2$.  Globular clusters that
spend their time nearer the Galactic center were not modeled by Vesperini
\& Heggie, although the general trend for such clusters is preferential
loss of low mass stars due to disk shocking and tidal stripping.  NGC 6121
has $R_{peri} \approx 1$ kpc (Peterson, Rees, \& Cudworth 1995) and it is
somewhat surprising that it still exists.  Nontheless, it does exist, and
its low mass stars exhibit a MF slope $\alpha \approx -2.3$ (Richer
1997).  For this PDMF slope the WD mass fraction should be little affected
by stellar evaporation.  I conclude that despite the probably large amount
of stellar evaporation this cluster has suffered, its kinematical
evolution should not have significantly altered the WD mass fraction.

Although counter-intuitive, stellar evaporation in open clusters may not
preferentially eject low mass stars since mass segregation spares the low
mass members many encounters, particularly with the oft-produced central
massive binary system (Terlevich 1987).  It is not yet clear, however,
what the relative evaporation of WDs versus the entire range of main
sequence stars is expected to be.  Weidemann \etal (1992) tried to address
this problem specifically for the Hyades.  They argued that extrapolation
of the Hyades PDMF up to $8 M_{\sun}$ would predict at least 21 more
cluster WDs than currently reside in the Hyades (seven single WDs and
three in binaries).  Additionally, the coolest of the known Hyades WDs has
a cooling age of 300 Myrs, about half the cluster age.  They argued that
all the missing WDs were the older ones, which have had time to escape,
and which perhaps had their velocities augmented by asymmetric mass loss
during planetary nebulae ejection or dissolution of their precursor
binary.  To address how the Hyades might dissolve they numerically
integrated test particles in representative Galactic orbits.  They
concluded that evaporation of light stars in the Hyades has reduced the
original population by a factor of perhaps ten.  While Weidemann \etal did
not specifically say how the Hyades WD mass fraction might evolve, their
numbers indicate that despite a near dissolution of the Hyades, the WD
mass fraction should not have changed by more than a factor of two.  Even
if $90$\% of the original Hyades stars have been lost, if they were
preferentially low mass members, less than $90$\% of the cluster mass
would have been lost.  This number compares closely with the $\geq 68$\%
fraction of WDs lost (currently $10$, formerly more than $31$).

Although the other three open clusters presented in Figures 2 and 3 have
not been individually treated by theoretical studies, some guidance can be
gained by the work of de la Feunte Marcos (1996).  His N-body open cluster
simulations disrupted after an average of $\sim115$ Myrs for N = $250$
particles.  The disruption time increased with the number of cluster
members.  All three of these open clusters are about 115 Myrs old (the
Pleiades is $70$ Myrs old, NGC 2168 is $85$ Myrs old, and NGC 3532 is
$165$ Myrs old), yet all three were born with significantly more than
$250$ stars (see Table 1).  These clusters not only still do exist, but
they should still exist, and their stellar evaporation losses should not
be catastrophic.  Thus, by analogy with the Hyades which seems to have
approximately retained its WD mass fraction despite stellar evaporation,
these three younger open clusters should be even less affected by stellar
evaporation.

In summary, for the particular clusters studied here, kinematical
evolution has been moderate (the Pleiades, NGC 2168, and NGC 3532) to
strong (the Hyades and NGC 6121). {\it Nevertheless, kinematical evolution
has little changed the WD mass fractions in these five clusters.  All five
clusters have approximately the WD mass fraction that would be produced by
a stellar populations with a Salpeter-like IMF.}  The insensitivity of the
WD mass fraction to the cluster dynamical history is due to the fact that
most WDs have masses intermediate between the top and bottom of the
present main sequence in every cluster.

\subsection{Implications for the Galactic Disk and Halo Field Populations}

How similar are the cluster WD mass fractions to those of the Galactic
disk and halo?  The essence of the question is how similar the field star
IMF is to that of the observed clusters.  For the open clusters and the
Galactic disk, the expectation is that the IMFs should be essentially the
same since current work on star-forming complexes (Hillenbrand \etal 1993;
Hillenbrand 1997), on open clusters (e.g. Reid 1992; von Hippel \etal
1996), and on the disk field population (e.g.\ Gould, Bahcall, \& Flynn
1997) all yield similar mass functions.  While the disk field population
includes stars of all ages, most studies of the Galactic star formation
history (e.g.\ Twarog 1980; Pardi \& Ferrini 1994) have concluded that the
rate of star formation in the disk has been falling somewhat with time.
Thus, the median stellar age of the disk is likely to be greater than half
the disk age.  Assuming a disk age of $\sim10$ Gyrs (Winget \etal 1987;
Liebert, Dahn, \& Monet 1988; Oswalt \etal 1996) then for a median disk
star age of $5$ to $7$ Gyrs and $\alpha = -2.35$ (see Table 3), the
Galactic disk WD mass fraction should be $6$ to $7$\%.  This number is
consistent with the empirical estimate of $3$ to $7$\% (Liebert, Dahn, \&
Monet 1988; Oswalt \etal 1996).

For the globular clusters and the Galactic halo the situation is much less
clear, even though a few globular cluster luminosity functions have now
been measured to luminosities equivalent to nearly $0.1 M_{\sun}$
(e.g.\ De Marchi \& Paresce 1995a,b; Elson \etal 1995; Cool, Piotto, \&
King 1996).  The greatest current difficulty is measuring the halo
luminosity function, which is presently poorly known and controversial.
Nontheless, the theoretical cluster simulations can again act as a guide.
Consistently, larger clusters and clusters with steeper IMFs survive
longer.  Thus, the halo field star population was likely produced by
clusters that were smaller and/or had a flatter IMF.  The halo field IMF
should not be too much flatter than the cluster IMFs however, or it would
violate a number of nucleosynthesis constraints (see Gibson \& Mould 1997,
and references therein).  Assuming the Galactic halo is $\sim12$ Gyrs old
(Reid 1997b; Gratton \etal 1997; Chaboyer \etal 1998), it should contain
$8$ to $9$\% ($\alpha = -2.35$) or perhaps as much as $15$ to $17$\%
($\alpha = -2.0$) of its stellar mass in the form of WDs (see Figure 2 and
Table 3).

Continued observations of globular clusters and the next generation of
combined N-body and stellar evolution models (Tout \etal 1997) should
refine both our estimates of the WD mass fraction and the relationship
between the cluster and the field star IMFs.  It would be of particular
interest to know if the disk and halo field star IMFs were in any way
different from the open and globular cluster IMFs (as opposed to the
PDMFs).  This would indicate whether the types of star clusters that we
find today are typical of the entire range of all star clusters ever
formed.  The WD mass fractions provide a particularly useful tool in this
work as they are less sensitive to cluster dynamics than the MFs.

\section{Conclusion}

I have undertaken a literature search through 31 July 1997 of white dwarfs
(WDs) in open and globular clusters.  I have tried to make a careful
evaluation in each case of the likelihood that the object is a white dwarf
and that it is a cluster member.  The results are presented for 13 open
clusters and 11 globular clusters.  Currently there are 36 single WDs and
5 WDs in binaries known among the open clusters, and 340 single WDs and 11
WDs in binaries known among the globular clusters.  From these data I have
calculated WD mass fractions for four open clusters (the Pleiades, NGC
2168, NGC 3532, and the Hyades) and one globular cluster (NGC 6121).  I
develop a simple model of cluster evolution that incorporates stellar
evolution but not dynamical evolution to interpret the WD mass fractions.
I augment the results of my simple model by turning to sophisticated
N-body simulations incorporating stellar evolution (Terlevich 1987; de la
Feunte Marcos 1996; Vesperini \& Heggie 1997).  I find that even though
these clusters undergo a range of degrees of kinematical evolution from
moderate (the Pleiades, NGC 2168, and NGC 3532) to strong (the Hyades, NGC
6121) the WD mass fraction is relatively insensitive to kinematical
evolution and little changed from a model incorporating only stellar
evolution with a Salpeter-like initial mass function.  By comparing the
cluster mass functions to that of the Galactic disk, and incorporating
plausibility arguments for the mass function of the Galactic halo, I
estimate the WD mass fraction in these two field populations.  I assume
the Galactic disk is $\sim10$ Gyrs old (Winget \etal 1987; Liebert, Dahn,
\& Monet 1988; Oswalt \etal 1996) and that the Galactic halo is $\sim12$
Gyrs old (Reid 1997b; Gratton \etal 1997; Chaboyer \etal 1998), although
the WD mass fraction is insensitive to age within this regime.  I find
that the Galactic halo should contain from $8$ to $9$\% ($\alpha = -2.35$)
or perhaps as much as $15$ to $17$\% ($\alpha = -2.0$) of its stellar mass
in the form of WDs.  The Galactic disk WD mass fraction should be $6$ to
$7$\% (for a median stellar age of $5$ to $7$ Gyrs and $\alpha = -2.35$),
consistent with the empirical estimates of $3$ to $7$\% (Liebert, Dahn, \&
Monet 1988; Oswalt \etal 1996).  Ultimately, precise comparisons between
the field and cluster MFs for both the disk and halo would be a means of
determining if the clusters we see today are typical of those that built
the field populations.

\acknowledgments

I would like to thank Chris Tout for help with his stellar evolution
parameterizations, Neill Reid for informative discussions of the Hyades,
and Ken Mighell for helpful comments on the manuscript.  Support for this
work was provided by NASA through grant number GO-6424 from the Space
Telescope Science Institute, which is operated by the Association of
Universities for Research in Astronomy, Inc., under NASA contract
NAS5-26555.  Support for this work was also provided by a grant from the
Edgar P. and Nona B.  McKinney Charitable Trust.  This research has made
use of NASA's Astrophysics Data System Abstract Service.


\eject

\newpage

\figcaption[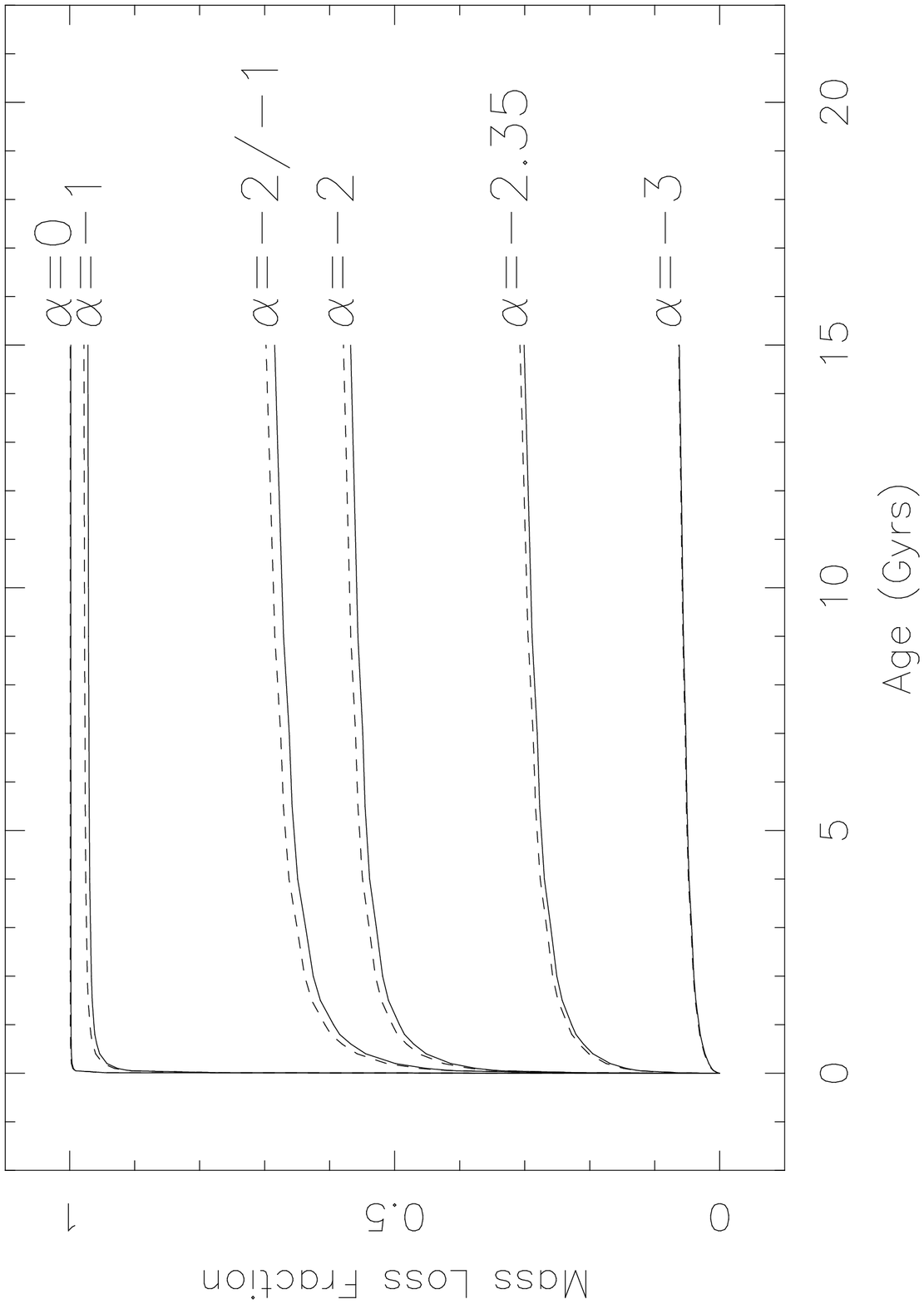]
{The fraction of mass lost from the model clusters as a function of age,
up to $15$ Gyrs, for IMFs characterized by slopes $\alpha = 0, -1, -2,
-2.35$, and $-3$, and a double power-law slope with $\alpha = -2$ above
$0.6 M_{\sun}$ and $\alpha = -1$ otherwise (as advocated Gould, Bahcall,
\& Flynn (1997) for the Galactic disk field stars).  On this system,
the slope of $-2.35$ is the Salpeter (1955) value.  The dashed lines are
for model runs with $M_{up} = 5 M_{\sun}$ and the solid lines are for
$M_{up} = 8 M_{\sun}$.}

\figcaption[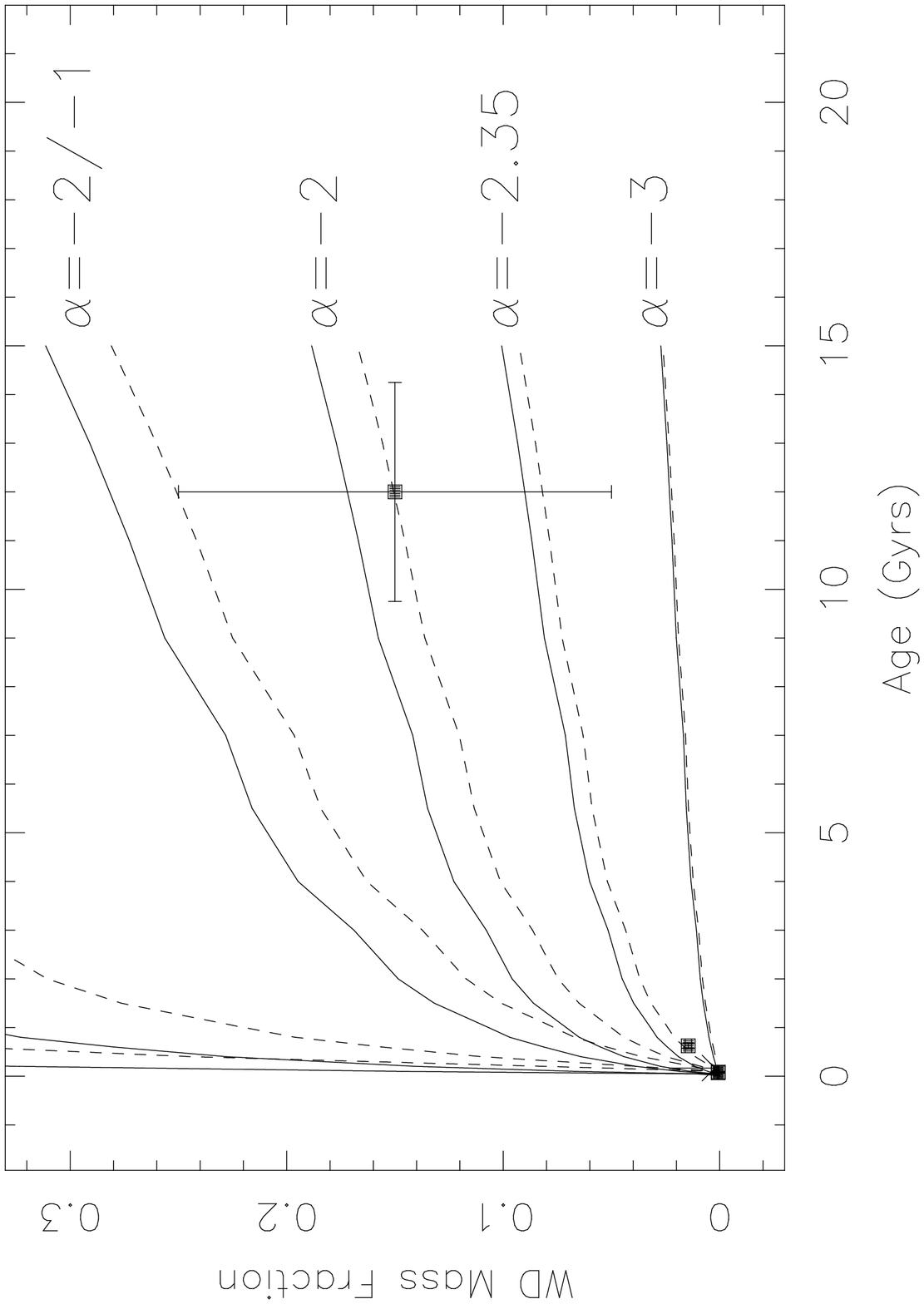]
{The WD mass fraction for the model clusters as a function of the IMF
slope, for $M_{up} = 5$ (dashed lines) and $8$ (solid lines) $M_{\sun}$.
Also plotted are the four open cluster and one globular cluster values,
along with their 1 $\sigma$ uncertainties.  The arrow symbol near Log(age)
= $-0.8$ Gyrs is the upper limit value for NGC 3532.  On this scale the
open cluster values are difficult to separate from the model lines near
the origin.}

\figcaption[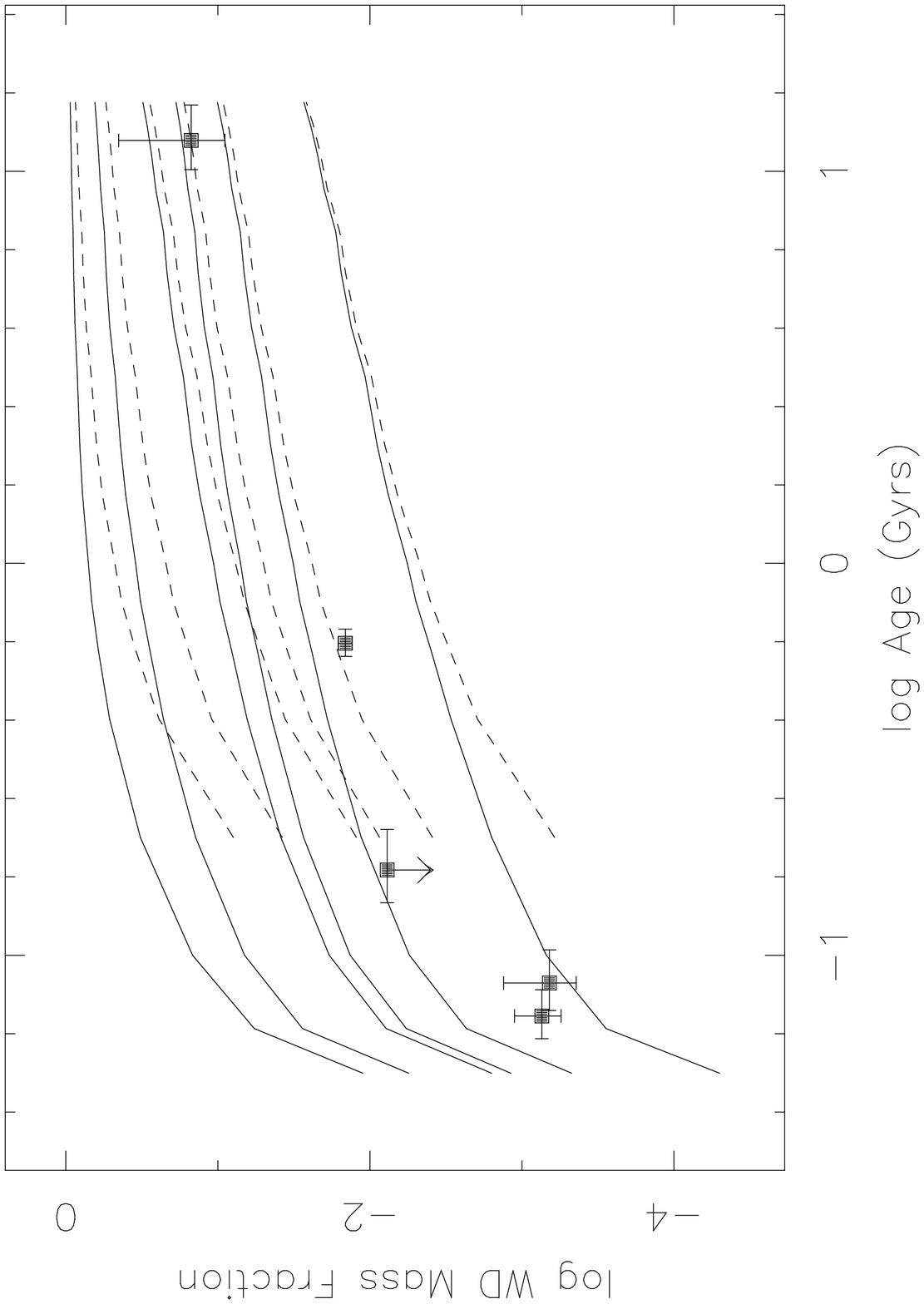]
{Similar to Figure 2 except that both axes are plotted as logarithms.  The
model runs are not labeled with the IMF slope for clarity of presentation,
but are, from top to bottom, $\alpha = 0, -1, -2/-1, -2, -2.35$, and
$-3$.}

\clearpage

\begin{deluxetable}{llrcccccclc}
\tablewidth{0pt}
\tablecaption{White Dwarfs in Open Clusters}
\tablehead{
\colhead{cluster} & \colhead{alias} & \colhead{$N_s$} & \colhead{ref}  &
\colhead{$N_b$}   & \colhead{ref}   & \colhead{$N_c$} & \colhead{Mass} &
\colhead{ref}     & \colhead{Age}   & \colhead{ref} \nl
\phantom{12}(1) & \phantom{1}(2) & (3) & (4) & (5) & (6) & (7) & (8) & (9) & (10) & (11)}
\startdata
Hyades   &     &  7 &  1,2    & 3 & 3,4   &  a  & 410-480                    &  5    &  0.63 &  6   \nl
Pleiades & M45 &  1 &  7,8,9  &   &       & 1-2 & 1000-2000                  & 10,11 &  0.07 & 12   \nl
NGC2168  & M35 &  2 &  7,13   &   &       &     & $\geq$1600-3200\phantom{1} & 14    &  0.09 & 7,13 \nl
NGC2287  & M41 &  2 &  8      &   &       &     &                            &       &  0.18 &  8   \nl
NGC2420  &     &  4 & 15      &   &       &     & $\geq$4000                 & 16    &  2.4  & 17   \nl
NGC2451  &     &  1 &  7,18   &   &       &     &                            &       &  0.07 & 18   \nl
NGC2477  &     &  4 & 15      &   &       &     &                            &       &  1.2  & 15   \nl
NGC2516  &     &  4 &  4      &   &       &     &                            &       &  0.14 & 19   \nl
NGC2632  & M44 &  4 & 20      &   &       &     &                            &       &  0.7  & 21   \nl
NGC2682  & M67 &  1 & 22      & 2 & 23,22 &     &                            &       &  4.0  & 19   \nl
NGC3532  &     &  6 & 7,24,25 &   &       &     & $\geq$600                  & 25    &  0.17 & 25   \nl
total    &     & 36 &         & 5 &       &     &                            &       &       &
\tablenotetext{}{NGC 2632 = Praesepe.}
\tablenotetext{a}{See discussion in Section 4.1.}
\tablerefs{
(1) Wegner, Reid, \& McMahan 1989; 
(2) Reid 1997a; 
(3) Boehm-Vitense 1993; 
(4) Koester \& Reimers 1996; 
(5) Reid 1992;
(6) Perryman \etal 1998;
(7) Reimers \& Koester 1988a; 
(8) Koester \& Reimers 1981; 
(9) Weidemann 1977; 
(10) Meusinger, Schilbach, \& Souchay 1996;
(11) van Leeuwen 1980;
(12) Stauffer, Hamilton, \& Probst 1994;
(13) Reimers \& Koester 1988b; 
(14) Leonard \& Merritt 1989;
(15) von Hippel, Gilmore, \& Jones 1995; 
(16) Leonard 1988;
(17) Demarque, Sarajedini, \& Guo 1994;
(18) Koester \& Reimers 1985; 
(19) Meynet, Mermilliod, \& Maeder 1993;
(20) Wagner \etal 1986;
(21) Mermilliod 1981;
(22) Pasquini, Belloni, \& Abbott 1994; 
(23) Landsman \etal 1997; 
(24) Koester \& Reimers 1993; 
(25) Reimers \& Koester 1989.}
\enddata
\end{deluxetable}

\clearpage

\begin{deluxetable}{llrcrcrrc}
\tablewidth{0pt}
\tablecaption{White Dwarfs in Globular Clusters}
\tablehead{
\colhead{cluster} & \colhead{alias} & \colhead{$N_s$} & \colhead{ref}  &
\colhead{$N_b$}   & \colhead{ref}   & \colhead{$N_c$} & \colhead{Mass} &
\colhead{ref}\nl
\phantom{12}(1) & \phantom{1}(2) & (3) & (4) & (5) & (6) & (7)\phantom{12} & (8)\phantom{123} & (9)}
\startdata
NGC104  & 47Tuc &   9 &  1 &  2 & 1,2 &        & 1,300,000 & 3      \nl
NGC5272 &  M3   &     &    &  1 &  4  &        &           &        \nl
NGC6121 &  M4   & 258 &  5 &    &     & 20,000 &    70,000 & 6,7,17 \nl
NGC6397 &       &  40 &  8 &  3 &  9  &        &           &        \nl
NGC6402 &  M14  &     &    &  1 & 10  &        &           &        \nl
NGC6539 &       &     &    &  1 & 11  &        &           &        \nl
NGC6624 &       &     &    &  1 & 12  &        &           &        \nl
NGC6752 &       &  21 & 13 &    &     &        &           &        \nl
NGC6838 &  M71  &  12 & 14 &    &     &        &           &        \nl
NGC7078 &  M15  &     &    &  1 & 15  &        &           &        \nl
Ter5    &       &     &    &  1 & 16  &        &           &        \nl
total   &       & 340 &    & 11 &     &        &           &
\tablerefs{
(1) Paresce, De Marchi, \& Jedrzejewski 1995;
(2) Ables \etal 1989;
(3) Meylan \& Mayor 1986;
(4) Hertz, Grindlay, \& Bailyn 1993;
(5) Richer \etal 1997;
(6) Richer \etal 1995;
(7) Sigurdsson 1993;
(8) Cool, Piotto, \& King 1996;
(9) Grindlay \etal 1995;
(10) Cote \etal 1997;
(11) D'Amico \etal 1993;
(12) Stella, White, \& Priedhorsky 1987;
(13) Renzini \etal 1996;
(14) Richer \& Fahlman 1988;
(15) Anderson \etal 1990;
(16) \'Ergma \& Fedorova 1991;
(17) Peterson, Rees, \& Cudworth 1995.}
\enddata
\end{deluxetable}

\begin{deluxetable}{rrrrr}
\tablewidth{0pt}
\tablecaption{White Dwarf Mass Fractions}
\tablehead{
\colhead{Age} & \colhead{$\alpha = -2$} & \colhead{$\alpha = -2.35$} 
              & \colhead{$\alpha = -2$} & \colhead{$\alpha = -2.35$} \nl
              & $M_{up} = 5$            & $M_{up} = 5$ 
	      & $M_{up} = 8$            & $M_{up} = 8$}
\startdata
 5 & 0.1094 & 0.0566 & 0.1309 & 0.0648 \nl
 7 & 0.1203 & 0.0631 & 0.1418 & 0.0713 \nl
10 & 0.1408 & 0.0757 & 0.1623 & 0.0839 \nl
12 & 0.1506 & 0.0818 & 0.1720 & 0.0900
\enddata
\end{deluxetable}

\end{document}